\documentclass[conference,a4paper]{IEEEtran}

\usepackage{cite}
\usepackage{amsmath,amssymb,amsfonts}
\usepackage{algorithmic}
\usepackage{graphicx}
\usepackage{textcomp}
\usepackage{xcolor}
\usepackage{booktabs}
\usepackage{multirow}
\usepackage{url}
\usepackage{eso-pic}
\usepackage[normalem]{ulem}

\title{AMAX: Accuracy-Aware Multi-Loop Asymmetric Perforation for Heterogeneous Architectures}

\author{
\IEEEauthorblockN{Amna Hassan, Paul Danielse, Anuj Pathania, Andy Pimentel, Semeen Rehman}
\IEEEauthorblockA{
University of Amsterdam\\
Amsterdam, Netherlands\\
\{a.hassan, p.e.g.danielse, a.pathania, a.d.pimentel, s.rehman\}@uva.nl
}
}

\begin{document}

\AddToShipoutPictureFG*{%
  \AtPageUpperLeft{%
    \put(0,-25){%
      \makebox[\paperwidth]{%
        \normalfont\small\itshape
        \uline{To appear in the Proceedings of the IEEE Nordic Circuits and Systems Conference (NorCAS), Tampere, Finland, 2026}%
      }%
    }%
  }%
}

\maketitle

\begin{abstract}
Thermal constraints in heterogeneous multi-core systems can significantly degrade performance, as conventional techniques such as dynamic voltage and frequency scaling (DVFS) and task migration introduce overhead under high utilization. Approximate computing, particularly loop perforation, can mitigate this overhead by reducing computation at the cost of accuracy loss. However, existing approaches typically apply a uniform perforation rate across all loops, ignoring loop-specific sensitivity. We propose {\em AMAX}, an accuracy-aware multi-loop asymmetric perforation framework that reduces the combinational design space through loop-wise profiling and multi-objective configuration selection, and dynamically adapts the resulting configurations at runtime according to performance requirements. Experiments on PARSEC benchmarks show that {\em AMAX} achieves up to 0.85 improvement in normalized accuracy over symmetric perforation for the same performance targets.
\end{abstract}


\section{Introduction}
\label{intro}

Thermal challenges in heterogeneous multi-core systems are aggravated by the unequal power densities of high-performance cores (P-cores) and energy-efficient cores (E-cores), leading to thermal hotspots and spatial thermal imbalances \cite{huang2022accurate,kim2020adaptive}. Conventional thermal management techniques, such as dynamic voltage and frequency scaling (DVFS) and task migration \cite{atienza202520}, can mitigate these effects but often introduce significant performance overhead. DVFS reduces frequency under thermal stress, while task migration depends on the availability of cooler cores, which becomes limited under high utilization.

Approximate computing can alleviate this overhead by reducing computation at the cost of accuracy loss. Loop perforation, an approximate computing technique, skips selected loop iterations to reduce workload, power consumption, and execution time. Existing approaches \cite{taufique2025exploiting,kanduri2018approximation} commonly use a single perforation rate across all loops. However, loops can differ substantially in their contribution to execution time and output accuracy, making uniform perforation suboptimal.

\begin{figure}
    \centering
    \includegraphics[width=0.5\textwidth]{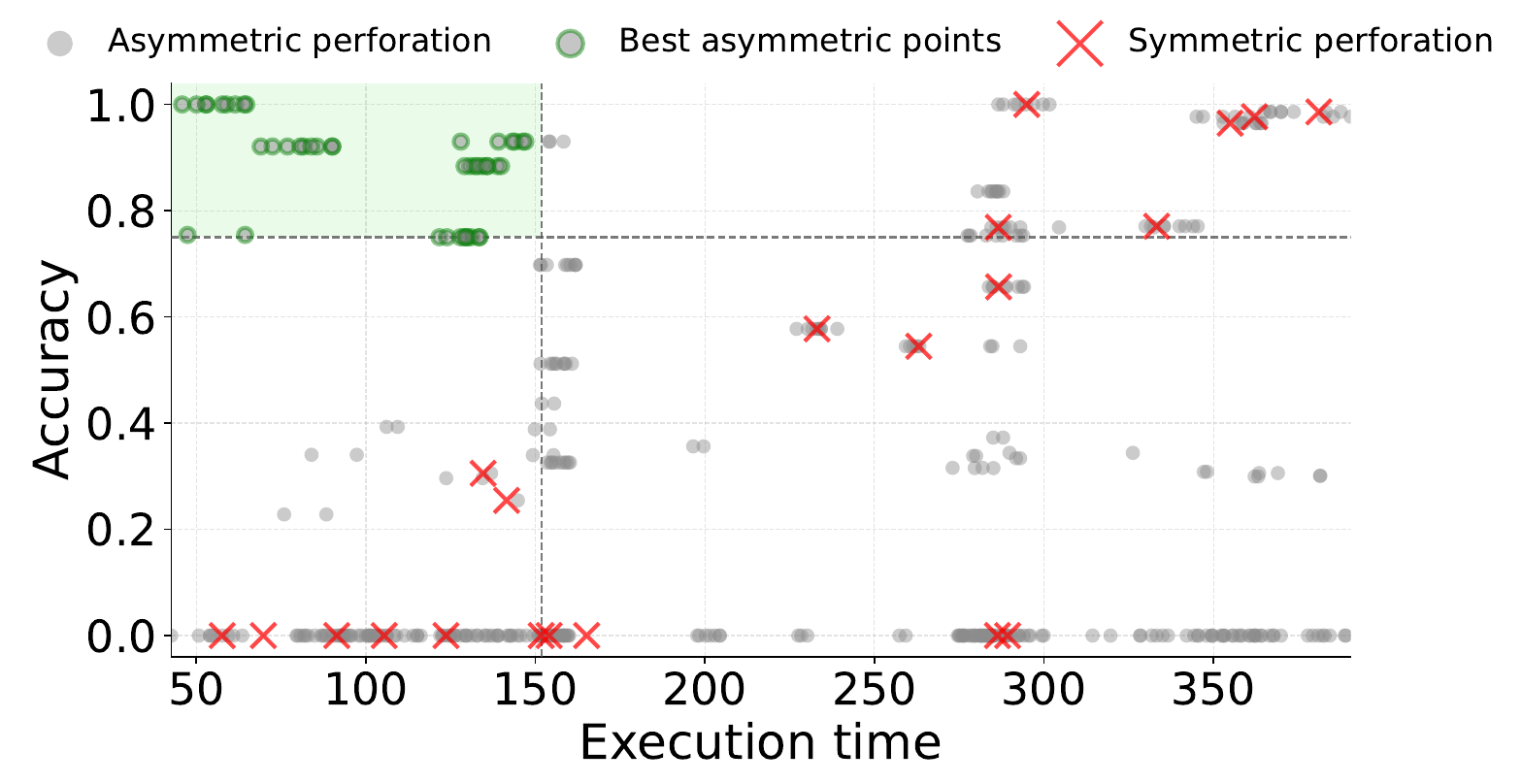}
    \caption{Performance-accuracy trade-offs of symmetric and asymmetric loop
perforation for Swaptions, with the shaded region indicating the desired
design space.}
    \label{fig:problem_identification}
\end{figure}
\begin{figure*}
    \centering
    \includegraphics[width=1\textwidth]{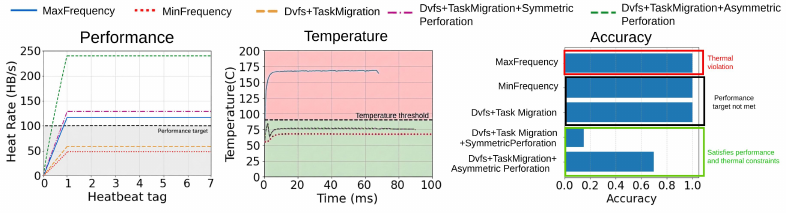}
    \caption{Performance, peak temperature, and accuracy of Swaptions on a heterogeneous multi-core system with four threads under different configurations.}
    \label{fig:motivation}
\end{figure*}

Fig.~\ref{fig:problem_identification} shows the performance--accuracy trade-off of symmetric and asymmetric loop perforation. Asymmetric perforation produces several high-accuracy, low-execution-time configurations, while symmetric perforation fails to achieve comparable trade-offs in the desired highlighted region. Uniformly perforating all loops either limits performance improvement or causes substantial accuracy degradation due to differences in loop sensitivity. By assigning independent rates to individual loops, asymmetric perforation achieves a more favorable balance between performance improvement and accuracy preservation. \par
Fig.~\ref{fig:motivation} illustrates the effect of thermal management and loop perforation on the Swaptions benchmark. Application performance is monitored using periodic heartbeats. A heartbeat indicates application progress at a predefined execution point, with each progress event identified by a heartbeat tag, while the heart rate $(HB/s)$ represents the number of progress events per second. Threads are mapped to heterogeneous cores under maximum and minimum frequency configurations, while thermal-aware DVFS and task migration are used to control thermal violations. Although these policies reduce thermal stress, they introduce substantial performance overhead. Loop perforation can recover performance, but symmetric and asymmetric strategies exhibit significantly different performance-accuracy trade-offs. Asymmetric perforation provides much higher performance while achieving an absolute improvement of 0.52 in normalized accuracy, motivating loop-specific perforation control. 
We therefore propose \textit{Accuracy-aware Multi-loop Asymmetric Perforation (AMAX)}, a framework that assigns loop-specific perforation rates through design-time profiling and dynamically selects configurations at runtime to provide fine-grained perforation control in thermally constrained heterogeneous systems. The primary contributions are summarized as follows:

\begin{itemize}
    \item \textbf{Multi-loop asymmetric perforation:} We demonstrate that uniform perforation is suboptimal because loops differ in their performance contribution and accuracy sensitivity, and introduce independent loop-specific perforation rates.

    \item \textbf{Sensitivity-guided multi-loop exploration:} We develop a loop-wise profiling method for \(L\) perforable loops and \(R\) candidate perforation rates per loop, requiring only \(O(LR)\) evaluations and providing an efficient basis for reducing the exponential \(O(R^L)\) multi-loop design space.

    \item \textbf{Runtime performance–accuracy adaptation:} We develop a runtime controller that selects and dynamically adjusts asymmetric loop configurations according to measured performance targets while avoiding unnecessary accuracy degradation.

    \item \textbf{HotSniper integration and evaluation:} We extend HotSniper, a widely used thermal simulation framework for multi-core systems, with dynamic multi-loop perforation support and instrument PARSEC benchmarks to evaluate {\em AMAX} under thermal and performance constraints.
\end{itemize}
\section{Related Work}
\label{relatedwork}

Loop perforation has been widely studied as a software-level approximation technique for improving performance while controlling output quality. \cite{misailovic2010exploring} investigated its impact on application accuracy, while \cite{misailovic2011probabilistic} analyzed how perforation errors depend on computational patterns. Adaptive approaches have also been proposed: \cite{carbin2013verified} activates perforation based on program integrity, \cite{hoffmann2009using} dynamically enables or disables perforated loops, and \cite{khudia2015rumba} detects large approximation errors at runtime
and selectively recomputes affected output elements. Loop perforation has further been extended to GPU workloads \cite{omar2017graphtuner,lashgar2018loop}, balancing performance gains against approximation error.

Most closely related to our work, Taufique et al. \cite{taufique2025exploiting} propose an approximation-aware performance-management framework for thermally constrained heterogeneous systems. It uses offline profiling to construct performance--accuracy models and dynamically adjusts approximation levels at runtime. However, approximation is treated as a single global control knob, such that perforable loops are governed by the same approximation level. Consequently, it neither exploits loop-specific sensitivity nor explores combinations of independently perforated loops.

In contrast, {\em AMAX} adaptively selects perforation rates for individual loops by exploiting differences in loop sensitivity. {\em AMAX} identifies optimal perforation configurations by using sensitivity-aware combinational exploration to provide finer performance--accuracy trade-offs.

 \begin{figure}[!t]
    \centering
    \includegraphics[width=1\columnwidth]{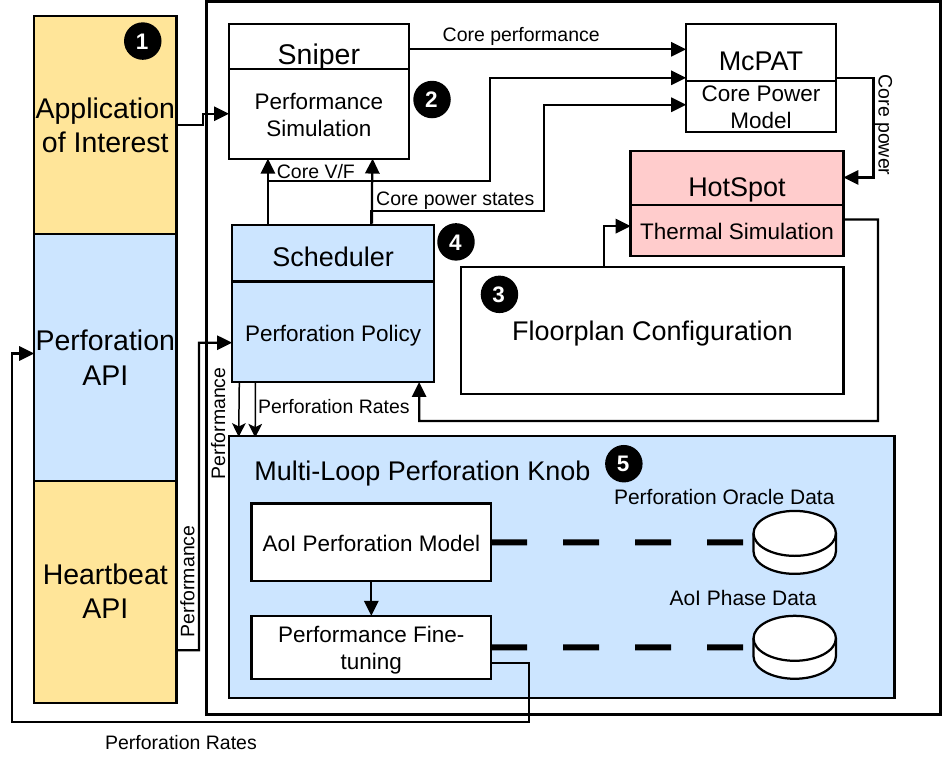}
    \caption{Integration of Multi-loop Perforation in HotSniper}
    \label{fig:architecture}
\end{figure}
 \begin{figure}[!t]
    \centering
    \includegraphics[width=1\columnwidth]{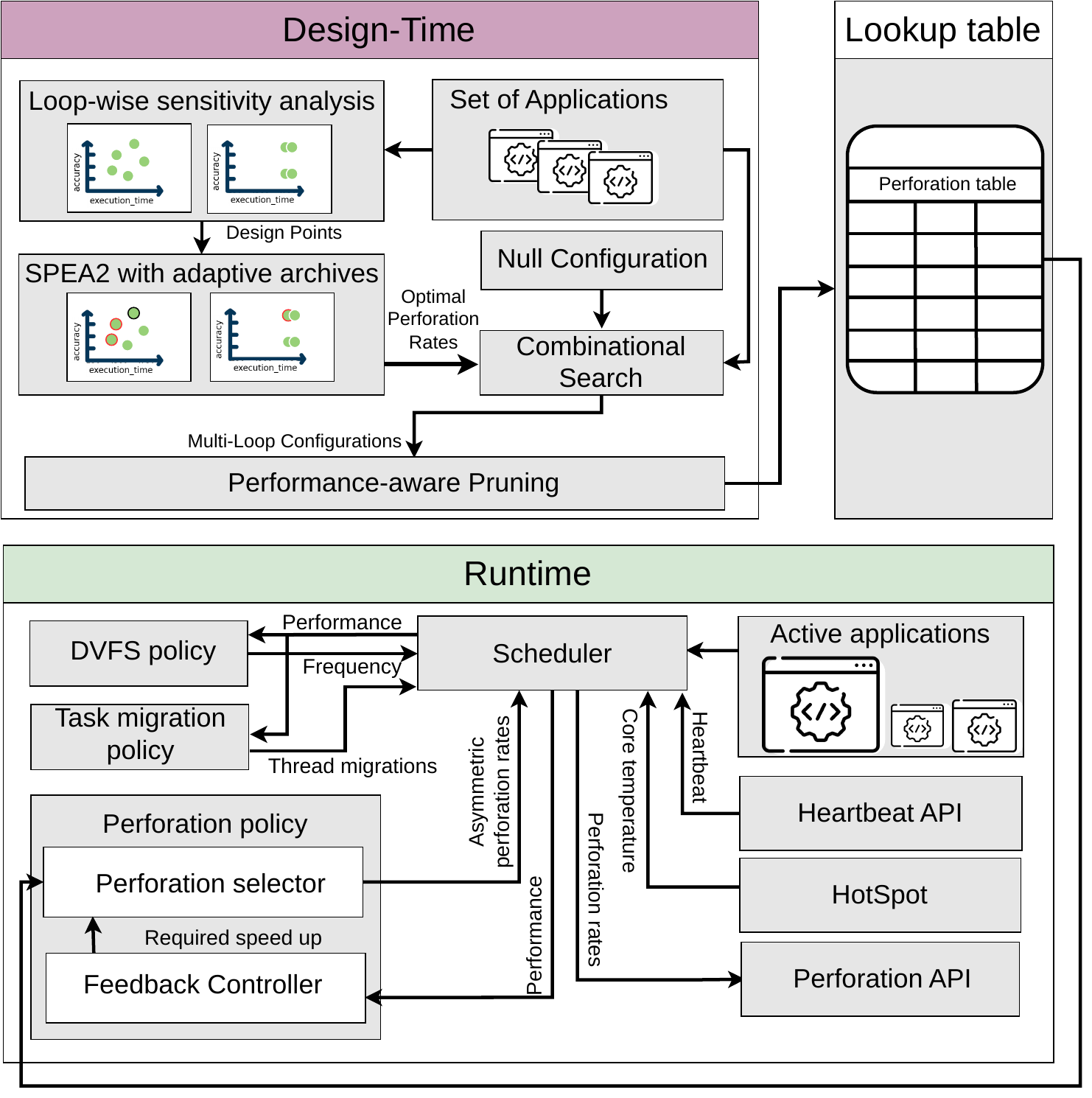}
    \caption{System overview of {\em AMAX} with design-time and runtime components}
    \label{fig:design_run}
\end{figure}
\section{Accuracy-Aware Multi-Loop Asymmetric Perforation}
\label{methodology}

{\em AMAX} consists of two main components. First, we extend HotSniper to support independent perforation rates for multiple loops. Second, {\em AMAX} combines design-time profiling and multi-objective optimization with a runtime feedback controller. At design time, loop-specific performance--accuracy trade-offs are identified and pruned. At runtime, the resulting configurations are dynamically selected according to performance and thermal constraints. Fig. ~\ref{fig:design_run} provides an overview of AMAX, showing its design-time profiling and runtime adaptation components.

\subsection{HotSniper Integration}

Fig.~\ref{fig:architecture} shows the integration of multi-loop perforation into HotSniper. The developed perforation library supports two execution modes. In native mode, loop-specific rates are provided through an environment variable (e.g., \texttt{MANUAL\_PERFORATION}). In HotSniper mode, the library is connected to the scheduler, enabling runtime adjustment of individual loop perforation rates.

The workflow is as follows. \textcircled{1} The application of interest (AoI) periodically generates heartbeats representing its performance, while a target performance is specified before execution. \textcircled{2} Sniper instruments and simulates the AoI using PinPlay \cite{patil2010pinplay}, an Intel framework to record and replay the application execution for simulation. \textcircled{3} The target multi-core floorplan used for thermal simulation. Performance and power information are used by the thermal model to estimate core temperatures. \textcircled{4} The scheduler uses this information to activate thermal-management policies and, when performance constraints are violated, adjust loop perforation rates. \textcircled{5} The multi-loop perforation knob exposes an application programming interface (API) for dynamically modifying individual loop rates and supports selective and truncation perforation. For each rate, an $N$-bit mask, where \(N\) is the number of iterations in the loop, determines which iterations are executed, supporting rates from 0\% to 100\%. The rates are retrieved from Sniper's statistics database and updated at designated execution points. Application accuracy is measured by comparing the perforated output with a reference execution.

\subsection{Offline Profiling and Configuration Selection}
\label{offline}

\subsubsection{Loop-wise Performance--Accuracy Sensitivity}

Consider an application $A$ with $L$ perforable loops and $R$ possible
perforation rates per loop. Exhaustive exploration requires evaluating
$R^L$ configurations and therefore becomes impractical as $L$ increases.

{\em AMAX} first performs a loop-wise sensitivity analysis by varying one loop
while keeping all others unperforated. For each loop
$i \in \{1,\dots,L\}$, configurations of the form
\(
(r_i,0,\dots,0)
\)
are evaluated, reducing the initial exploration complexity from
\(
O(R^L)\) to \(O(LR).
\)
The resulting execution time and output error quantify each loop's
sensitivity and identify loops that provide high performance gains with
limited accuracy degradation. This analysis reduces
the candidate space before jointly evaluating multiple perforated loops.

\begin{figure*}
\centering
\includegraphics[width=1\textwidth]{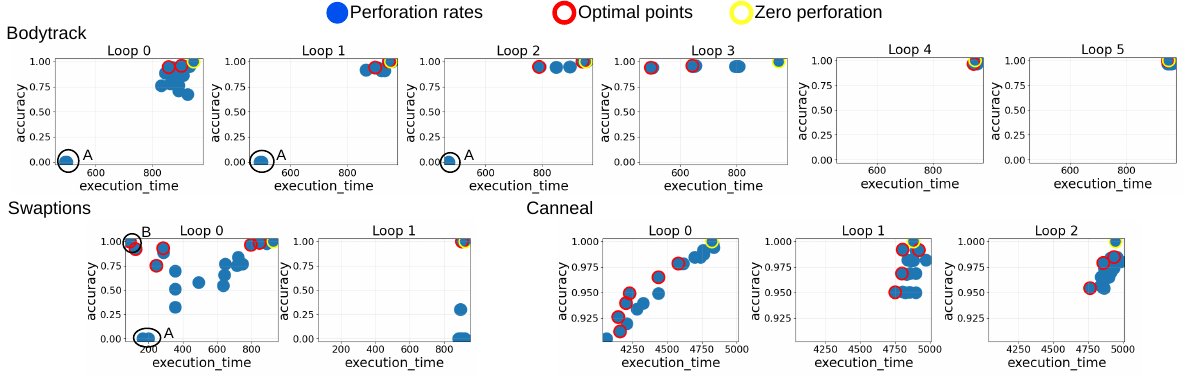}
\caption{Per-loop sensitivity and selected configurations in the
accuracy--execution-time design space.}
\label{fig:designspace}
\end{figure*}

\subsubsection{SPEA2-based Configuration Selection}

The sensitivity analysis is formulated as a bi-objective optimization
problem, minimizing execution time and output error. For loop $i$, let
\(
X_i=\{x_1,x_2,\dots,x_R\},
\)
where $x_j$ represents the $j$th perforation rate applied to loop $i$.
Each configuration is evaluated as
\[
f_1(x_j^i)=T(x_j^i), \qquad
f_2(x_j^i)=E(x_j^i),
\]
where $T(\cdot)$ and $E(\cdot)$ denote execution time and output error,
respectively.

We use SPEA2 \cite{zitzler2001spea2}, a multi-objective genetic algorithm, to retain a diverse set of
performance--accuracy trade-offs. Unlike simple Pareto filtering, which
may retain many closely spaced non-dominated configurations, SPEA2 ranks
points using both dominance and density. This enables {\em AMAX} to retain a
compact and diverse archive, directly reducing the subsequent multi-loop
combinational search space. A configuration $x_j^i$ dominates $x_k^i$
when
\[
T(x_j^i) \leq T(x_k^i)-0.05T(x_o),
\qquad
E(x_j^i) \leq E(x_k^i),
\]
with at least one strict inequality, where $x_o$ denotes execution without
perforation. We denote this dominance relation by $x_j^i \prec x_k^i$.

SPEA2 assigns each configuration a strength
\begin{equation}
S(x_j^i)=
\left|
\left\{
x_k^* \in P \mid x_j^i \prec x_k^*
\right\}
\right|,
\label{eq:strength}
\end{equation}
where \(P\) denotes the current population of candidate configurations, a raw fitness
\begin{equation}
R(x_j^i)=
\sum_{x_k^* \in P,\;x_k^*\prec x_j^i}
S(x_k^*),
\label{eq:rawfit}
\end{equation}
and a density term
\begin{equation}
D(x_j^i)=\frac{1}{\sigma_i^k+2},
\label{eq:density}
\end{equation}
is computed using the $k$-th nearest neighbor distance $\sigma_i^k$
resulting in
\begin{equation}
F(x_j^i)=R(x_j^i)+D(x_j^i).
\label{eq:fit}
\end{equation}

\paragraph{Adaptive Archive Size}

Unlike standard SPEA2, {\em AMAX} adjusts the archive size according to each
loop's performance sensitivity. For loop $l$,
\[
\Delta T_l=T_l^{\max}-T_l^{\min},
\]
and its archive size is
\[
\bar{A}_l=
A_{\min}
+\alpha
\frac{\Delta T_l}{\max_m \Delta T_m},
\]
where $A_{\min}$ is the minimum archive size and $\alpha$ controls scaling and $\max_m \Delta T_m$ is the maximum time spread across all loops.
Consequently, loops with larger performance variation retain more
configurations, while less sensitive loops require fewer. The $|A_i|$
configurations with the lowest fitness are retained for loop $i$.

\subsubsection{Combinational Search and Final Pruning}

The configurations retained in each loop archive, together with a
zero-perforation option, form a reduced combinational design space.
If loop $i$ retains $|A_i|$ candidates, the joint search contains
\[
N_{\mathrm{comb}} =
\prod_{i=1}^{L} (|A_i|+1)
\]
configurations, compared with $R^L$ in exhaustive exploration. Thus,
the overall design-time exploration requires
\[
O\left(LR + \prod_{i=1}^{L}(|A_i|+1)\right)
\]
configuration evaluations.

This is necessary because independently optimal per-loop
configurations may not remain optimal when combined. {\em AMAX} therefore
evaluates only combinations of retained candidates, capturing cross-loop
interactions without exploring the full $R^L$ space. Configurations with
similar performance are then grouped, and only the highest-accuracy
configuration is retained. These resulting configurations are stored in an offline profile table during runtime selection.

Fig.~\ref{fig:designspace} illustrates the process. Region A contains
configurations that cause application failure and are discarded, while
Region B contains configurations that dominate alternatives in execution
time and accuracy.
\begin{figure}
    \centering
    \includegraphics[width=1.0\columnwidth]{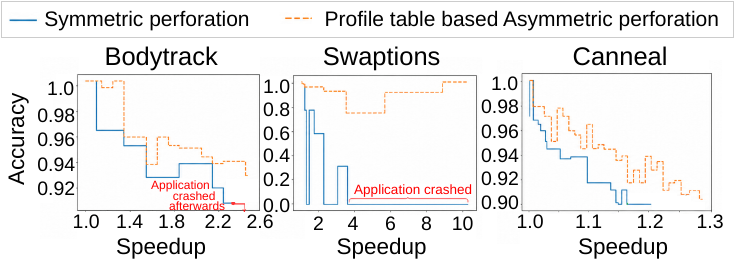}
    \caption{Accuracy and speedup of symmetric and asymmetric loop perforation for Bodytrack, Swaptions, and Canneal.}
    \label{fig:symvsasym}
\end{figure}

\subsection{Runtime Policies}

\subsubsection{DVFS Policy}

{\em AMAX} controls frequency independently for each core. Cores initially operate at minimum frequency and are scaled only when their power consumption remains below the assigned TDP:
\[
f_t=
\frac{\mathrm{TDP}(1-\beta)}{p}f_{t-1},
\]
where $p$ is the measured core power and $\beta$ is a safety margin. This enables localized power management while avoiding unnecessary performance degradation.

\subsubsection{Task Migration Policy}

We adopt the migration policy of Taufique et al. \cite{taufique2025exploiting}. When the power budget is exceeded, applications are migrated from P-cores to available E-cores if the expected performance loss is acceptable. Applications are migrated back to P-cores when their performance requirements cannot be satisfied and sufficient power headroom becomes available.
\subsubsection{Asymmetric Perforation Policy}
Our perforation policy consists of two components, i.e., an offline perforation selector and an online feedback controller. The performance target specifies the desired application performance level and is represented as a target speedup relative to the non-perforated execution. \par
The perforation selector provides an initial set of asymmetric perforation rates using the previously computed offline profile table. These perforation rates are selected based on the system's performance requirements, specifically the target speedup. It begins by selecting the two neighboring points around the target speedup: (i) the closest configuration whose speedup is less than or equal to the target, and (ii) the nearest configuration whose speedup exceeds the target. Then it considers all configurations with speedup greater than the target and chooses the one that achieves the highest accuracy among them. The accuracy of this selected higher-speedup configuration is compared with the accuracy of the lower-speedup configuration. If the lower-speedup configuration provides higher accuracy and its speedup is within a small tolerance of the target, it is selected. Otherwise, the higher-speedup configuration is chosen. If the target lies outside the range of profiled configurations, then the nearest boundary configuration is selected. In this way, the selector favors accuracy while remaining close to the requested runtime speed. \par
The online feedback controller periodically measures achieved speedup and computes its deviation from the target. When the measured performance falls below the target by more than a predefined tolerance, the controller shifts to a more aggressive perforation configuration based on the new speedup requirement, but when the achieved speedup exceeds the target by a greater magnitude, the controller shifts to a less aggressive configuration to recover accuracy. Afterwards, the new selected configurations are applied by updating the perforation rates of all the loops.
\section{Implementation, Evaluation and Results}
\label{results}
 \begin{figure*}
    \centering
    \includegraphics[width=1.0\textwidth]{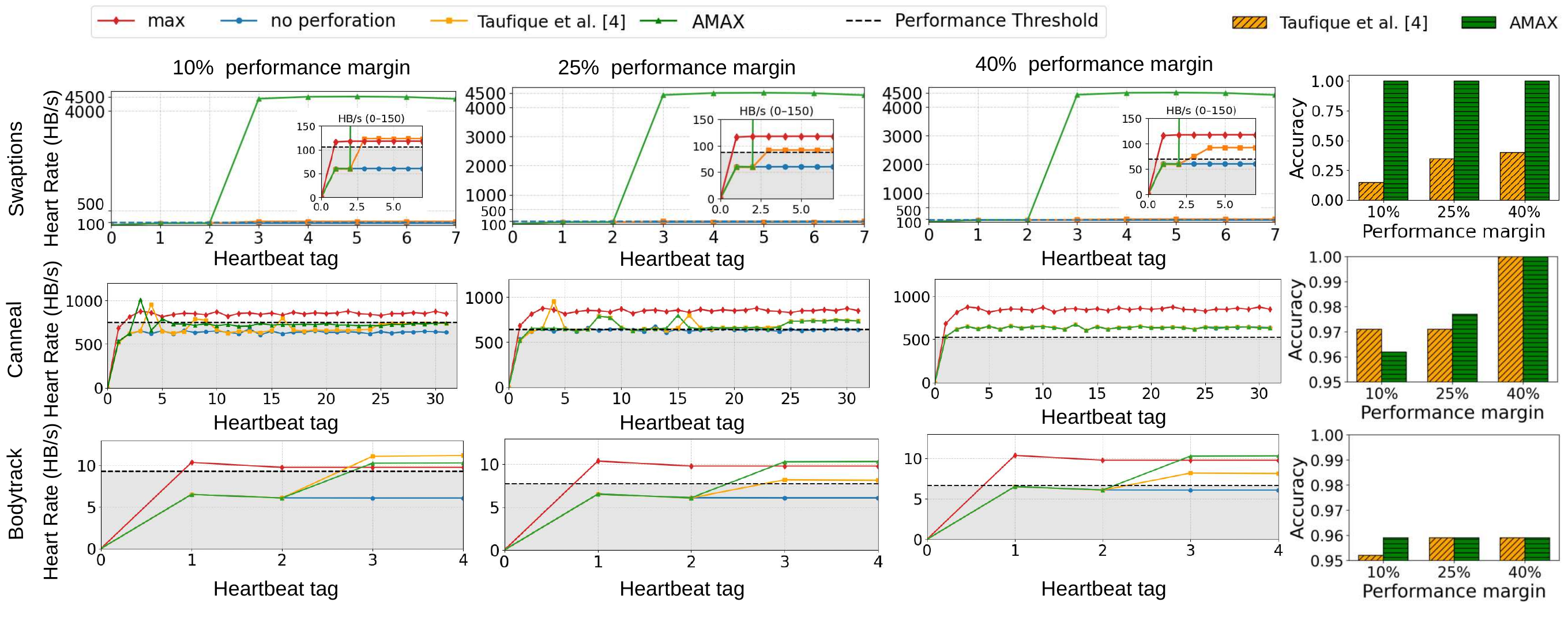}
    \caption{Runtime performance of {\em AMAX} under different performance targets on Bodytrack (simlarge), Swaptions (simsmall), and Canneal (simsmall), compared with the approach of  Taufique et al. \cite{taufique2025exploiting}, maximum performance, and no-perforation execution}
    \label{fig:results}
\end{figure*}
We evaluate {\em AMAX} on a heterogeneous multi-core architecture modeled using HotSniper \cite{pathania2018hotsniper}. We simulate 16 cores (8 P-cores and 8 E-cores), each with private L1/L2 caches. P-cores have 32KB instruction/64KB data L1 and 2MB L2 caches, while E-cores have 64KB instruction/32KB data L1 and 1MB L2 caches. All cores share a unified 4MB L3 cache. P-cores operate at 3.2--5.1 GHz and E-cores at 2.4--3.9 GHz. We implemented loop-specific dynamic perforation across six PARSEC benchmarks: Blackscholes \cite{black1973pricing}, Swaptions \cite{heath1992bond}, Canneal \cite{banerjee1994parallel}, Streamcluster \cite{o2002high}, Bodytrack \cite{deutscher2005articulated}, and x264 \cite{wiegand2003overview}. Blackscholes, Streamcluster and x264 exhibit little or no accuracy degradation under perforation, while the remaining applications show more distinct performance-accuracy trade-offs. We therefore focus on Swaptions (compute-intensive), Canneal (memory-intensive), and Bodytrack (compute- and memory-intensive) for detailed evaluation.

Output quality uses relative difference for Swaptions, Blackscholes, and Canneal, vector difference for Bodytrack, Peak Signal-to-Noise Ratio (PSNR) and output file size for x264, and clustering accuracy for Streamcluster. Accuracy is normalized to the non-perforated reference, where 1.0 denotes the reference output and lower values indicate larger deviations. Benchmarks contain 1–6 perforable loops in computation-intensive regions targeted for execution-time reduction.

Heartbeats monitor application performance and adjust perforation rates to meet targets. The heartbeat range spans minimum- and maximum-performance scenarios, with the maximum obtained by mapping threads to P-cores at the highest frequency. Targets are 10\%, 25\%, and 40\% below maximum performance.
\par

\begin{figure}
    \centering
    \includegraphics[width=1.0\columnwidth]{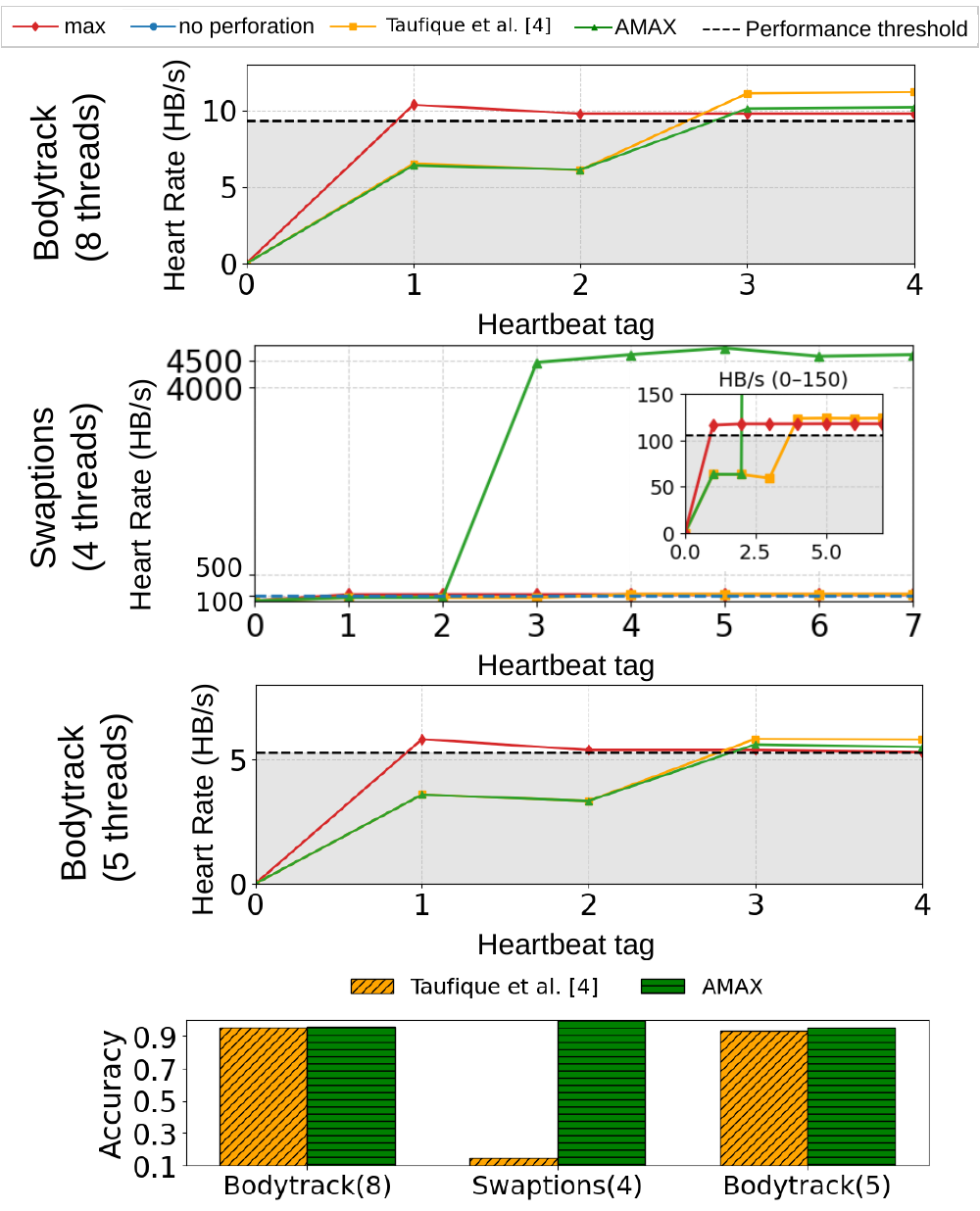}
    \caption{Runtime performance and accuracy of {\em AMAX} and \cite{taufique2025exploiting} on simultaneous execution of heterogeneous workloads, including Bodytrack (8 threads), Swaptions (simsmall), and Bodytrack (5 threads).}
    \label{fig:results_multi}
\end{figure}

To ensure measurement stability, profiling experiments are conducted in isolation on the same DAS-6 \cite{bal2016medium} node with identical CPUs at fixed frequency. Each design point is executed 20 times, and we report the average execution time with the coefficient of variation (CoV)
\[
\text{CoV} = \frac{\sigma}{\mu},
\]
where \( \mu \) and \( \sigma \) denote the mean and standard deviation. Only measurements with CoV below 0.05 are considered valid. \par

Fig. \ref{fig:symvsasym} compares symmetric and asymmetric perforation, with asymmetric configurations obtained through the design-time profiling in Section \ref{offline}. Asymmetric perforation achieves higher accuracy for the same performance gain, while symmetric perforation is more likely to cause application failure under aggressive perforation, as shown for Swaptions and Bodytrack. For Canneal, asymmetric perforation achieves additional performance gains because loops contribute unequally to execution time, allowing optimization to target performance-critical loops more effectively. \par

{\em AMAX} uses DVFS and task migration, following Taufique et al. \cite{taufique2025exploiting}, to enforce thermal constraints. Asymmetric perforation complements DVFS and task migration by recovering performance lost to thermal management with minimal accuracy degradation. Thus, under the same maximum-temperature constraint, {\em AMAX} improves the performance--accuracy trade-off without altering thermal management.

Fig. \ref{fig:results} presents the runtime performance of {\em AMAX} compared with Taufique et al. \cite{taufique2025exploiting}, maximum-performance, and no-perforation execution. Swaptions and Canneal use simsmall input to provide multiple data points, whereas Bodytrack uses simlarge because simsmall contains only one frame. Bodytrack and Canneal are evaluated with 8 threads and Swaptions with 4 threads, across targets of 10\%, 25\%, and 40\% below maximum performance. For Swaptions, {\em AMAX} achieves significant performance gains without accuracy loss, consistently selecting the dominant configuration in region B of Fig. \ref{fig:designspace} by prioritizing the highest-accuracy configuration satisfying the target. Overall, {\em AMAX} outperforms Taufique et al. \cite{taufique2025exploiting} across all targets, achieving up to 0.85 higher normalized accuracy. \par

For Canneal, at a 10\% performance margin, the target is unrealistic because Fig.~\ref{fig:symvsasym} shows limited performance gain from perforation. Therefore, {\em AMAX} uses the highest-performance configuration to approach the target without prioritizing accuracy. At 25\%, it tracks the target to preserve accuracy, while at 40\%, perforation is unnecessary because the target is already met without approximation. For Bodytrack, variation between targets is small because maximum performance is close to the no-perforation baseline. Nevertheless, Fig.~\ref{fig:symvsasym} shows that perforation can provide additional performance gains. Similar to Swaptions, the selector chooses a dominant configuration within a local window. {\em AMAX} therefore meets targets while avoiding unnecessary accuracy degradation, whereas Taufique et al. \cite{taufique2025exploiting} trades accuracy more aggressively for performance. \par

Similar behavior is observed in Fig. \ref{fig:results_multi} during simultaneous execution of two Bodytrack instances (8 threads,5 threads) and one Swaptions instance (4 threads). The runtime algorithm maintains performance while preserving accuracy, with {\em AMAX} achieving higher accuracy across all three applications. Thus, for fixed accuracy, {\em AMAX} achieves higher performance, while for the same performance gain, it achieves higher accuracy. Accuracy degradation occurs only when required to satisfy the target, demonstrating the effectiveness of offline profiling and adaptive runtime control.
\section{Conclusion}
\label{conclusion}We proposed {\em AMAX}, an accuracy-aware framework for multi-loop asymmetric perforation in thermally constrained heterogeneous systems. Unlike traditional symmetric perforation approaches, {\em AMAX} utilizes loop-level sensitivity analysis and multi-objective optimization to identify efficient performance–accuracy trade-offs at design time, and dynamically adapts perforation rates at runtime to meet performance targets. Evaluation on PARSEC benchmarks shows that {\em AMAX} achieves higher accuracy compared to prior approaches while satisfying runtime performance constraints. {\em AMAX} effectively exploits loop heterogeneity to avoid unnecessary accuracy loss and improve performance efficiency. Furthermore, we integrated multi-loop perforation into the HotSniper framework, enabling fine-grained, loop-specific control of perforation rates. This allows both scheduler-level and external control of loop-specific perforation rates, providing greater flexibility for evaluating performance–accuracy trade-offs.

\bibliographystyle{IEEEtran}
\bibliography{references}

\end{document}